\documentclass[journal]{IEEEtran}
\usepackage[table]{xcolor}
\usepackage[english]{babel}
\usepackage[utf8]{inputenc}
\usepackage{eurosym}
\usepackage{float}
\usepackage{flushend}
\usepackage{setspace}
\usepackage{amssymb}
\usepackage{tikz}
\usetikzlibrary{fit}
\usetikzlibrary{backgrounds}
\usetikzlibrary{positioning,fit,calc}
\usepackage{bbding}
\usepackage[]{tabularx}
\usepackage{circuitikz}

\usepackage{booktabs,multirow}

\usepackage{moreverb}
\usepackage{enumerate}
\usepackage{textcomp}
\makeatletter
\let\MYcaption\@makecaption
\makeatother
\usepackage[font=footnotesize]{subcaption}
\makeatletter
\let\@makecaption\MYcaption
\makeatother
\usepackage{graphicx}
\usepackage{amssymb}
\usepackage{amsthm}
\usepackage{amsmath}
\usepackage{multirow}
\usepackage{array}
\usepackage{bm}
\usetikzlibrary{arrows}
\usepackage{enumitem}
\usepackage{arydshln}
\usepackage{mathtools}
\newcommand\myhat[1]{\smash{\hat{#1}}\vphantom{#1}}
\newcommand\mybreve[1]{\smash{\breve{#1}}\vphantom{#1}}

\begin{document}

\title{Geometric Algebra Power Theory in Time Domain}
	
\author{Francisco Gil Montoya, Raul Ba\~{n}os, Alfredo Alcayde, Francisco Arrabal-Campos, \\ and   Javier Rold\'{a}n-P\'{e}rez, {\em Member, IEEE}}
\maketitle

\allowdisplaybreaks

\begin{abstract} 
In this paper, the power flow in electrical systems is modeled in the time domain by using geometric algebra and the Hilbert transform. The use of this mathematical framework overcomes some of the limitations shown by the existing methodologies under distorted supply or unbalanced load. In that cases, the instantaneous active current may not be the lowest RMS current for all the possible conditions. Moreover, this current can contain higher levels of harmonic distortion compared to the supply voltage and it cannot be applied to single phase systems. The proposed method can be used for sinusoidal and non-sinusoidal power supplies, non-linear loads, single- and multi-phase systems, and it provides meaningful engineering results with a compact formulation. Several examples have been included to verify the validity of the proposed theory.
\end{abstract}    
\section{Introduction}
\label{sec.introduction}
Pioneer power theories for analysing electrical systems were developed by Steinmetz, Kennelly and Heaviside, among others, by the end of the XIX century~\cite{steinmetz1893complex,kennelly1893impedance,heaviside1892electrical}. Nowadays, these theories are still a source of discussion and debate concerning their correctness and physical interpretation~\cite{czarnecki2005instantaneous}. Some of them were formulated in the frequency domain, such as those proposed by Budeanu~\cite{budeanu1927puissances} and Czarnecki~\cite{czarnecki2008currents}, while other ones where formulated in the time domain, like those presented by Fryze~\cite{fryze1932wirk}, Akagi~\cite{akagi1984instantaneous} and Depenbrock~\cite{depenbrock1993fbd}.  More recently, Lev-Ari~\cite{lev2006decomposition} and Salmer\'{o}n~\cite{salmeron2009instantaneous} have made relevant contributions to the field by using the Hilbert Transform (HT) and tensor algebra, respectively. All these theories are devoted to explain the power-transfer process between complex electrical systems and they establish mathematical concepts associated to fictitious powers (e.g. reactive power), which are of a great value from the engineering point of view. Unfortunately, none of the existing proposals can be used to separate current components in the time domain under any type of voltage distortion, asymmetry, non-linearity of the load or combinations thereof. Some of these limitations have been reported in the literature~\cite{czarnecki2005instantaneous,haley2015limitations}. 

In this paper, a new proposal is presented to overcome these limitations by applying two mathematical tools: geometric algebra (GA) and the HT. GA is a versatile tool that can be used to model different physical and mathematical problems~\cite{hestenes2012clifford}. The application of GA makes it possible to separate current components that have engineering relevance for systems with any number of phases (including single-phase systems)~\cite{montoya2020analysis,lev2009instantaneous}. The term ``engineering relevance'' was explicitly used while the term ``physical relevance'' was avoided, mainly because one of the main applications of power theories is current decomposition for load compensation purposes. Currently, there is still controversy regarding the physical meaning of these currents, although it is clear that they are relevant for the engineering practice. Current decomposition is addressed from both instantaneous and averaged points of view so that currents that do not contribute to active power can be compensated. On the other hand, thanks to the use of the HT \cite{saitou2002generalized,nowomiejski1981generalized}, the proposed theory can be applied to single-phase systems seamlessly, which is a relevant advantage compared to the existing theories. Mathematical demonstrations are omitted in this paper for the sake of conciseness. The paper includes an overview of GA in order to make the paper self-contained. However, for detailed information about GA and its applications to electrical systems, see~\cite{Montoya2020revisiting,chappell2014geometric}.
\section{Geometric Algebra and Power Theory}
\subsection{Hilbert Transform and Geometric Algebra Fundamentals}
Consider a multi-phase system where $u(t)$ is the voltage and $i(t)$ is the current. Both signals are arrays of multiple dimensions associated to a Hilbert space. Currents and voltages are periodic and $T$ is the period value. 
The HT ($\bm{\mathcal{H}}$) defined in time domain is the convolution of the Hilbert transformer $1/\pi t$ and a function $f(t)$:
\begin{equation}
\bm{\mathcal{H}}\left[f(t)\right]
=
\frac{1}{\pi}PV\int_{-\infty}^{+\infty}\frac{f(\tau)}{\tau-t}d\tau
\end{equation}
\noindent where $PV$ is the Cauchy principal value to handle the singularity at $t=\tau$. The Bracewell criteria is used to select the sign of the transformation~\cite{bracewell1986fourier}. The HT is a crucial tool to calculate quadrature signals and impedances in the time domain~\cite{lev2006decomposition}. In fact, it can be demonstrated that HT delays positive frequency components (fundamental and harmonics) by ${\pi}/{2}$~\cite{saitou2002generalized}. In general, $\|\bm{\mathcal{H}}\left[f(t)\right]\| \neq \|f(t)\|$ due to the presence of a \textit{dc} component~\cite{kschischang2006hilbert}. By using the HT, voltages and currents in electrical circuits are usually transformed to analytic functions in the complex domain:
\begin{flalign}
\vec{u}(t)
&=
u(t)+j \bm{\mathcal{H}}\left[u(t)\right] \\
\vec{i}(t)
&=
i(t)+j \bm{\mathcal{H}}\left[i(t)\right] \end{flalign}
Note that the use of a complex algebra is not a necessary condition and other algebras can be used as well. 

%Revisar lo de instanteo o no
%which can be also used to define the so-called instantaneous impedance~\cite{lev2006decomposition}. 
%"Una de las claves del buen desempeño de la teoría es el uso de la transformada de Hilbert" Ojo si pones algo así, nos pueden dar caña.
%
Consider now an orthonormal base $\bm{\sigma}=\{ \bm{\sigma}_1,\bm{\sigma}_2,\ldots,\bm{\sigma}_n\}$ defined for a vector space in $\mathcal{R}^n$. Then, it is possible to establish a new geometric vector space $\mathcal{G}^n$ with a bilinear operation. Under these assumptions, a vector can be represented as:
\begin{equation}
	\bm{a}=\sum_{n}a_n\bm{\sigma}_n=a_1\bm{\sigma}_1+\ldots+a_n\bm{\sigma}_n
\end{equation}
Typically, the coefficients $a_n$ are real numbers, but they could be complex numbers or even vectors from another $\mathcal{G}^m$ space.
In this new space, the geometric product between two vectors ($\bm{a}$ and $\bm{b}$) can be defined as:
\begin{equation}
\bm{M}
=
\bm{ab}
=
\bm{a}\cdot\bm{b}+\bm{a}\wedge\bm{b}
\end{equation}
which can be seen as the sum of the traditional scalar (or inner) product plus the so-called wedge (or Grassmann) product~\cite{hestenes2012clifford}. The latter fulfils the following property:
\begin{equation}
	\bm{a}\wedge\bm{b}=-\bm{b}\wedge\bm{a}
	\label{ec.bivector}
\end{equation}
This geometrical entity is commonly known as {\em bivector} and it cannot be found in traditional linear algebra~\cite{hestenes2012clifford}.  
%De esta forma, es posible definir la parte conmutativa y anticommutativa del producto geométrico como
%
%\begin{align}
%\bm{a}\cdot\bm{b} &= \frac{1}{2}\left(\bm{ab}+\bm{ba} \right)=\frac{1}{2}\left(\bm{M}+\bm{M}^\dagger\right)  \\
%\bm{a}\wedge\bm{b}&= \frac{1}{2} \left(\bm{ab}-\bm{ba} \right) =\frac{1}{2}\left(\bm{M}-\bm{M}^\dagger\right)
%\label{eq:com_anticom}
%\end{align}
%
%siendo $\bm{M}^\dagger$ el reverse de $\bm{M}$.
%
\subsection{Geometric Power in Time Domain}
\subsubsection{Linear loads}
For a general $n$-phase electrical system, an array containing the instantaneous voltages can be defined as follows:
\begin{equation}
{u}(t)
=
\left[u_1(t), u_2(t), \: \ldots \:, u_n(t)\right]
\end{equation}
where each voltage is referred to a virtual star point that might not be the same as the neutral conductor. An array representing node injected currents is also defined:
\begin{equation}
{i}(t)
=
\left[i_1(t), i_2(t),\: \ldots \:, i_n(t)\right]
\end{equation}
In order to simplify the notation, $u_k(t)=u_k$ and $i_k(t)=i_k$. 
The expression for the instantaneous power consumed by the circuit is widely known, and it can be calculated as:
\begin{equation}
	p(t)=u_1i_1+u_2i_2+\ldots+u_ni_n
	\label{eq:pot_instant}
\end{equation}
which represents the inner product between $u(t)$ and $i(t)$.
%y así ha sido reconocido por diferentes autores que han trabajado profusamente en la materia. %\cite{akagi1984instantaneous, willems1992new, salmeron2009instantaneous}. 

%The use of GA allows a spatial approach of voltage and current vectors. 
For a linear load, ${u}(t)$ and ${i}(t)$ can be represented in the GA domain as the geometric vectors $\bm{u}$ and $\bm{i}$, respectively, by means of a $2n$ dimensional space with base vectors  $\bm{\sigma}=\{\bm{\sigma}_{1},\bm{\sigma}_{\smash{\hat{1}}},\bm{\sigma}_{2},\bm{\sigma}_{\myhat{2}},\ldots,\bm{\sigma}_{n},\bm{\sigma}_{\myhat{n}}\}$, where $n$ is the number of phases:
%La aplicación de algebra geométrica permite realizar un enfoque espacial para los vectores de tensión y corriente. 
%Conjuntamente con la transformada de Hilbert, los  vectores de tensión y corriente geométricos se pueden expresar como
%
\begin{equation}
	\setlength{\arraycolsep}{0pt}
	\begin{array}{ r *{5}{ >{{}}c<{{}} r } }
		\bm{u} &=&\dfrac{1}{\sqrt{2}}(
		u_1\bm{\sigma}_1 &+&
		\bm{\mathcal{H}}\left[u_1\right]\bm{\sigma}_{\myhat{1}} &+&
		\cdots &+&
		u_n\bm{\sigma}_{n} &+&
		\bm{\mathcal{H}}\left[u_n\right]\bm{\sigma}_{\myhat{n}})
		\\[2ex]
		\bm{i} &=&\dfrac{1}{\sqrt{2}}(
		i_1\bm{\sigma}_1 &+&
		\bm{\mathcal{H}}\left[i_1\right]\bm{\sigma}_{\myhat{1}} &+&
		\cdots &+&
		i_n\bm{\sigma}_{n} &+&
		\bm{\mathcal{H}}\left[i_n\right]\bm{\sigma}_{\myhat{n}})
	\end{array}
	\label{eq:volt_curr_vector} 
\end{equation}
%

%The negative signs have been included so that an inductive element leads to a positive value of reactive power consumption, while a capacitive element leads to a negative one. 
The above expressions represent multidimensional analytic signals in the GA domain. The use of the HT in (\ref{eq:volt_curr_vector}) becomes essential to overcome the shortcomings of some existing time domain power theories \cite{czarnecki2005instantaneous,haley2015limitations,de2005discussion} and it is one of the main contributions of this work. These contributions are strongly supported by previous works of Nowomiejski, Saitou and Lev-Ari~\cite{nowomiejski1981generalized,saitou2002generalized,lev2006decomposition}. Note that the use of the HT is mandatory for single-phase calculations, but it can be omitted in the case of systems with more than one phase provided that an instantaneous approach is used, as in the $IRP$ theory. For this particular case, the basis is $n$ dimensional $\bm{\sigma}=\{\bm{\sigma}_{1},\bm{\sigma}_{2},\ldots,\bm{\sigma}_{n}\}$ and the current and voltage becomes:

\begin{equation}
	\setlength{\arraycolsep}{0pt}
	\begin{array}{ r *{5}{ >{{}}c<{{}} r } }
		\bm{u} &=&\dfrac{1}{\sqrt{2}}(
		u_1\bm{\sigma}_1 &+&
		\cdots &+&
		u_n\bm{\sigma}_{n} )
		\\[2ex]
		\bm{i} &=&\dfrac{1}{\sqrt{2}}(
		i_1\bm{\sigma}_1 &+&
		\cdots &+&
		i_n\bm{\sigma}_{n} )
	\end{array}
	\label{eq:volt_curr_vector} 
\end{equation}

For the most general case, the norms of the geometric voltage and current vectors are:
\begin{equation}
	\begin{aligned}
		\|\bm{u}\|=\sqrt{\bm{u}\bm{u}}=\sqrt{\bm{u}\cdot\bm{u}}&=\sqrt{\sum_{k=1}^{n} u_k^2+\bm{\mathcal{H}}^2\left[u_k\right]}\\
		\|\bm{i}\|=\sqrt{\bm{i}\bm{i}}=\sqrt{\bm{i}\cdot\bm{i}}&=\sqrt{\sum_{k=1}^{n} i_k^2+\bm{\mathcal{H}}^2\left[i_k\right]}
	\end{aligned}
\end{equation}

The instantaneous geometric power is defined as the product of voltage and current vectors, thus
\begin{equation}
	\bm{M}=\bm{ui}=\bm{u}\cdot\bm{i}+\bm{u}\wedge\bm{i}
	\label{eq:geom_power_time}
\end{equation}
%The inspection of (\ref{eq:geom_power_time}) reveals the potential of GA applied to electrical circuits. 
Similar expressions have been obtained in the literature by using other modelling tools (complex numbers, matrix algebra, vector calculus, quaternions, tensors, etc.). However, this one is more compact and unifies the application of GA tools. This expression consists of two terms that have a different nature, i.e., a scalar and a bivector term. This mathematical entity is called a multivector and it can be written as
\begin{equation}
	\bm{M}=M_p+\bm{M}_q
	\label{eq:mp_mq}
\end{equation}
where
\begin{align}
	M_p&=\bm{u}\cdot \bm{i}=\dfrac{1}{2}\sum_{k=1}^{n} u_{k} i_{k} + \bm{\mathcal{H}}\left[u_{k}\right] \bm{\mathcal{H}}\left[i_{k}\right] \nonumber \\
	&=\dfrac{1}{2}(u_1i_1+\bm{\mathcal{H}}\left[u_1\right]\bm{\mathcal{H}}\left[i_1\right]+\ldots+u_n i_n+\bm{\mathcal{H}}\left[u_{n}\right]\bm{\mathcal{H}}\left[i_{n}\right] \nonumber) \\[0.5cm]
	\bm{M}_q&=\bm{u}\wedge \bm{i} = 
	\setlength{\arraycolsep}{3pt}
	\dfrac{1}{2}\begin{vmatrix}
		\bm{\sigma}_1 &  \bm{\sigma}_{\myhat{1}} &  \ldots & \bm{\sigma}_{n}  & \bm{\sigma}_{\myhat{n}} \\[0.3cm]
		u_1 & \bm{\mathcal{H}}\left[u_{1}\right] & \ldots & u_{n} & \bm{\mathcal{H}}\left[u_{n}\right] \\[0.3cm] 
		i_1 & \bm{\mathcal{H}}\left[i_{1}\right] & \ldots & i_{n} & \bm{\mathcal{H}}\left[i_{n}\right]
	\end{vmatrix}  = \\[2ex] \nonumber
	&=\dfrac{1}{2}[\left(u_1\bm{\mathcal{H}}\left[i_{1}\right]-\bm{\mathcal{H}}\left[u_{1}\right]i_1\right)\bm{\sigma}_{1\myhat{1}}+\left(u_1 i_{2}-u_{2} i_1\right)\bm{\sigma}_{12} \\[1ex] \nonumber
	&+ \ldots + \left(u_n \bm{\mathcal{H}}\left[i_{n}\right]-\bm{\mathcal{H}}\left[u_{n}\right] i_n\right)\bm{\sigma}_{n\myhat{n}}]
	\label{eq:Mq}
\end{align}

The term $M_p$ is the scalar part and it includes the instantaneous active power $p(t)$. It will be referred to as {\em parallel geometric power}. The term $\bm{M}_q$ is the bivector part and will be named quadrature geometric power. It comprises the well-known \textit{instantaneous reactive power} in the \textit{IRP} theory and its further refinements~\cite{salmeron2009instantaneous, AkagiInstantaneous, willems1992new,dai2004generalized}. In (\ref{eq:Mq}), the determinant can be calculated by using Leibniz or Laplace formula.

%la potencia geométrica paralela, incluye a la potencia instantánea $p(t)$. Por otro lado,  $\bm{M_q}$, la potencia geométrica en cuadratura,   incluye  a  la conocida \textit{potencia reactiva instantanea} en la teoría $IRP$ y sus posteriores actualizaciones \cite{AkagiInstantaneous, willems1992new,dai2004generalized, salmeron2009instantaneous}.

The instantaneous geometric power $\bm{M}$ can also be written in terms of commutative and anti-commutative parts:
%La descomposición de la potencia geométrica instantánea también puede ser hallada como la parte conmutativa y anticonmutativa de $\bm{M}$
%
\begin{align}
	M_p &= \frac{1}{2}\left(\bm{ui}+\bm{iu} \right)=\frac{1}{2}\left(\bm{M}+\bm{M}^\dagger\right)  \\
	\bm{M_q}&= \frac{1}{2} \left(\bm{ui}-\bm{iu} \right) =\frac{1}{2}\left(\bm{M}-\bm{M}^\dagger\right)
	\label{eq:com_anticom}
\end{align}
\noindent where $\bm{M}^\dagger$ is the reverse of the instantaneous geometric power. It is worth noting that no matrices nor tensors are used in the definitions of powers presented in (\ref{eq:geom_power_time})$-$(\ref{eq:com_anticom}). This leads to a compact formulation that simplifies mathematical expressions.
%
%donde $\bm{M}^\dagger$ es el reverse de la potencia geométrica. Un detalle importante que merece la pena resaltar es que en las definiciones introducidas en (\ref{eq:geom_power_time})$-$(\ref{eq:com_anticom}) no se hace uso de matrices ni tensores. 
%
%Esto simplifica bastante los cálculos a realizar.
%
\subsubsection{Non-linear loads}
If the load is nonlinear, the current $i(t)$ has additional harmonics that are not present in the voltage source. These harmonics are included in $i_{\perp}(t)$ and some authors refers to it as ``out-of-band'' current~\cite{lev2006decomposition}. The total current can then be expressed as:
\begin{equation}
i(t)
=
i_{\parallel}(t)
+
i_{\perp}(t)
\end{equation}
where 
\begin{equation}
\begin{array}{ll}
i_{\parallel}(t)&=\left[i_{1_{\parallel}}(t), i_{2_{\parallel}}(t),\: \ldots \:, i_{n_{\parallel}}(t)\right]\\
i_{\perp}(t)&=\left[i_{1_{\perp}}(t), i_{2_{\perp}}(t),\: \ldots \:, i_{n_{\perp}}(t)\right]\\
\end{array}
\end{equation}
Therefore, the geometric transformation should include this new current component.
%De esta forma, la transformación geométrica debe incluir a esta nueva componente de la corriente. 
To this end, the dimension of the space must be increased by a factor equal to the total number of phases, so the basis becomes
%De forma que la base ha aumentado en dimensionalidad debido a la incorporación de
$\bm{\sigma}=\{\bm{\sigma}_{1},\bm{\sigma}_{\myhat{1}},\bm{\sigma}_{\mybreve{1}},\bm{\sigma}_{2},\ldots,\bm{\sigma}_{\mybreve{n}}\}$. 
Therefore, for the most general case, a basis of $3n$ dimensions is required, where $n$ is the number of phases. Note that if the $dc$ term (if any) is present in the current but not in the voltage, it must be included in $i_{\perp}(t)$. %If the \textit{dc} component is present, then $3n+1$ dimensions will suffice. 
%En total, de manera general se necesita una base de dimensión $3n$, siendo $n$ el número de fases.
%Para ello, basta simplemente con aumentar la dimensionalidad del espacio en número igual al número de fases para acomodar dichos componentes. 
%
Then, the current and voltage expressions for a non-linear circuit becomes:
%La expresion para tensión y corriente en un circuito no lineal es la siguiente:
%
\begin{equation}
\begin{array}{ll}
\bm{u} &=\dfrac{1}{\sqrt{2}}(
u_1\bm{\sigma}_1 +
\bm{\mathcal{H}}\left[u_1\right]\bm{\sigma}_{\myhat{1}} +
\cdots 
+u_n\bm{\sigma}_{n} +
\bm{\mathcal{H}}\left[u_n\right]\bm{\sigma}_{\myhat{n}})
\\[1ex]
\bm{i} &=\dfrac{1}{\sqrt{2}}(
i_{1_{\parallel}}\bm{\sigma}_1 +
\bm{\mathcal{H}}\left[i_{1_{\parallel}}\right]\bm{\sigma}_{\myhat{1}} +
i_{1_{\perp}}\bm{\sigma}_{\mybreve{1}} +\ldots \\[1ex]
&+i_{n_{\parallel}}\bm{\sigma}_n +
\bm{\mathcal{H}}\left[i_{n_{\parallel}}\right]\bm{\sigma}_{\myhat{n}} +
i_{n_{\perp}}\bm{\sigma}_{\mybreve{n}} )
=\bm{i}_{\parallel}+\bm{i}_{\perp}
\end{array}
\label{eq:volt_curr_vector2} 
\end{equation}
\noindent where $i_{k_{\parallel}}$ and $i_{k_{\perp}}$ are the component of the $k$-th harmonic current that is present and not present in the voltage, respectively.
In this case, the geometric apparent power becomes:
%La potencia aparatente geométrica en este caso es
%
\begin{equation}
\begin{aligned}
 \bm{M}&=\bm{ui}=\bm{u}(\bm{i}_{\parallel}+\bm{i}_{\perp})\\
 &=\bm{M_{\parallel}}+\bm{M_{\perp}}=M_p+\bm{M}_q+\bm{M}_{\perp}
\end{aligned}
\label{eq:mp_mq_nl}
\end{equation}
It can be seen that the power consist of an ``in-band'' and  ``out-of-band'' geometric power.  
%De forma que la potencia total está compuesta por una potencia geométrica "in-band" y una potencia geométrica "out-of-band". 
The former  includes the terms already present in (\ref{eq:mp_mq}).
%La potencia "in-band" incluye los términos ya presentados en (\ref{eq:mp_mq}).
%
\subsection{Current Decomposition}
\label{sec:current_decomposition}
If $i(t)$ is the instantaneous current demanded by a load, it can be separated into meaningful engineering components by manipulating~(\ref{eq:geom_power_time}) or (\ref{eq:mp_mq_nl}) (depending on the  nature of the load). For a linear load, left multiplying by the inverse of the voltage vector leads to:
%
%Efectivamente, operando por la izquierda con el inverso de la tensión, se tiene
%
\begin{equation}
	\bm{u^{-1}}\bm{M}=\bm{u^{-1}}\bm{ui}=\bm{i}
	\label{eq:v_inverse_en_M}
\end{equation}
since $\bm{u^{-1}\bm{u}}=1$. 
%ya que el producto $\bm{u^{-1}\bm{u}}=1$. 

The inverse of a vector in the geometric domain is:
\begin{equation}
	\bm{u^{-1}}=\frac{\bm{u}}{\|\bm{u}\|^2}
\end{equation}
which can be used in (\ref{eq:v_inverse_en_M}) to find the current decomposition. By replacing  (\ref{eq:mp_mq}) in (\ref{eq:v_inverse_en_M}):
\begin{equation}
	\begin{aligned}
		\bm{i}&=\bm{u^{-1}M}=\frac{\bm{u}}{\|\bm{u}\|^2}\bm{M}=\frac{\bm{u}}{\|\bm{u}\|^2}\left(M_p+\bm{M}_q\right) \\
		&=\frac{\bm{u}}{\|\bm{u}\|^2}M_p+\frac{\bm{u}}{\|\bm{u}\|^2}\bm{M}_q=\bm{i}_p+\bm{i}_q
	\end{aligned} 
	\label{eq:current_decomposition_initial}
\end{equation}

%
%where $\frac{M_p}{\|\bm{u}\|^2}$ is a time-dependent proportional constant between the current and the voltage. 
The above expression resembles that of Shepherd and Zakikhani \cite{shepherd1972suggested}. The term $\bm{i}_p$ is commonly known as \textit{instantaneous geometric parallel current}, while the term
%
%El término $\bm{i_p}$ es la denominada \textit{corriente paralela instantánea} y es el resultado proyectar la corriente $\bm{i}$ en la tensión $\bm{u}$, 
%siendo $\frac{M_p}{\|\bm{u}\|^2}$ una constante (dependiente del tiempo) de proporcionalidad entre dicha tensión y corriente.
$\bm{i}_q$ is the~\textit{instantaneous geometric quadrature current} and is orthogonal to $\bm{i}_p$.
%y representa la parte de la corriente instantánea ortogonal a $\bm{i_p}$. 
It should be noted that $\bm{i}_q$ is not a pure \textit{reactive current} since it is not always related to energy oscillations in reactive elements like inductors or capacitors~\cite{de2010ac}. In fact, it includes the effects generated by asymmetries of voltage sources and load unbalance.
%Esta corriente no recibe el nombre de \textit{corriente reactiva} ya que, de manera general, no se corresponde con ningún fenómeno físico relacionado con variación de energía en elementos reactivos almacenadores  como bobinas o condensadores. 

%Incluye, además, los efectos de la asimetría en la tensión de alimentación y el desbalanceo en la carga.

The geometric Fryze current can also be defined by using GA:
%Es posible acomodar también la corriente de Fryze definida por
%
\begin{equation}
	\bm{i}_F=\frac{\bar{M}_p}{\|\bm{\bar{\bm{u}}}\|^2}\bm{u}
\end{equation}
%
%siendo $\bar{M}_p$ el valor medio de la potencia geométrica paralela y $\|\bar{\bm{u}}\|$ el valor eficaz de la tensión geométrica. 
where $\bar{M}_p$ is the mean value of the geometric parallel power and $\|\bar{\bm{u}}\|^2$ is the RMS value of the squared norm of the geometric voltage.
It can be readily demonstrated that $\bar{M}_p=P$, where $P$ is the active power. Moreover, the reactive current defined by Budeanu and supported by Willems~\cite{willems2010budeanu}, Lev-Ari~\cite{lev2006decomposition} and Jeltsema~\cite{jeltsema2015budeanu} (among others) is:
%Es fácil demostrar que $\bar{M}_p=2P$, siendo $P$ la potencia activa. Asímismo, la intesidad reactiva de Budeanu apoyada por Willems \cite{willems2010budeanu}, Lev-Ari \cite{lev2006decomposition} o Jeltsema \cite{jeltsema2015budeanu} también puede hallarse como
%
\begin{equation}
	\bm{i}_B=\frac{\bar{M}_q}{\|\bm{\bar{\bm{u}}}\|^2}\bm{\mathcal{H}}\left[\bm{u}\right]
\end{equation}
where $\bar{M}_q$ is the mean value of the quadrature geometric power.
%siendo $\bar{M}_q$ el valor medio de la potencia geométrica en cuadratura. 
Similarly, $\bar{M}_q=Q$, where $Q$ is the reactive power defined by Budeanu.
%De forma similar a $\bar{M}_p$,  $\bar{M}_q=2Q$, siendo $Q$ la potencia reactiva de Budeanu.
Therefore, the current expression can be fully decomposed as follows:
%De esta manera, la descomposición  completa de la corriente  sería
%
\begin{equation}
	\bm{i}=\bm{i}_p+\bm{i}_q=\bm{i}_F+\bm{i}_f+\bm{i}_B+\bm{i}_b
	\label{eq:todas_corrientes}
\end{equation}
where $\bm{i}_f$ is the Fryze complementary current required to conform the parallel current. Similarly, $\bm{i}_b$ is the Budeanu complementary current required to conform the quadrature current. The asymmetry or unbalance current is included in~$\bm{i}_f$ and $\bm{i}_b$.

In the case of a non-linear load, only $\bm{M}_{\parallel}$ should be used in (\ref{eq:current_decomposition_initial}) for current decomposition purposes. The total current becomes
\begin{equation}
\begin{aligned}
	\bm{i}&=\bm{i}_{\parallel}+\bm{i}_{\perp}=\bm{i}_p+\bm{i}_q+\bm{i}_{\perp}\\
	&=\bm{i}_F+\bm{i}_f+\bm{i}_B+\bm{i}_b+\bm{i}_{\perp}
\end{aligned}
\label{eq:todas_corrientes2}
\end{equation}
%
%$\bm{i_f}$ es la corriente complementaria a la de  Fryze para conformar la corriente paralela. 
%Por otro lado, la corriente $\bm{i_b}$ es la complementaria de la de Budeanu para conformar la corriente en cuadratura. 
%La corriente de asimetria o desbalanceo está incluida en $\bm{i_b}$. 
%
The use of GA allows a natural decomposition of currents without requiring additional tools such as Clarke or Park transformations. The proposed methodology can be seamlessly applied to any distorted system since no constraints have been imposed to the waveforms of voltages and currents in this regard.
%Moreover, since no constraints have been imposed on the voltage and current, there are no restrictions to use the methodology of this paper in any distorted scenario. 
This theory can be applied to any circuit with an arbitrary number of phases, including single-phase systems. This cannot be achieved by using other theories such as the $IRP$, and it is a relevant feature of this proposal.
%El uso del álgebra geométrica permite la obtención de una descomposición más natural de la corriente sin  acudir a transformaciones adicionales como la de Clarke. Además, puesto que no se ha impuesto ninguna condición sobre la forma de onda de la tensión y la corriente, no existen restricciones para usar la metodología de este paper en cualquier situación de distorsión. Es evidente que esta teoría se puede aplicar a cualquier número de fases $n$, incluyendo los sistemas monofásicos, algo que no se ha podido realizar hasta la fecha con la teoría $IRP$.

%La transformación inversa al dominio original sería
The transformation from the geometric to the time domain for any current component in $\bm{i}_{\parallel}$ is:
\begin{equation}
	i(t) = \sqrt{2}\sum_{k=1}^{n}[\bm{i}_{\parallel}]_{{k}}
	\label{eq:projection_currents}
\end{equation}
\noindent where $[\cdot]_k$ refers to the $k$-th term of the geometric vector of the parallel current $\bm{i}_{\parallel}$.
%siendo $[\cdot]_k$ la componente $k$ del vector geométrico de corriente $\bm{i}_x$.
%
\section{Properties of the geometric power and current}
The instantaneous geometric power $\bm{M}$ and the instantaneous geometric current $\bm{i}$ fulfil a number of mathematical  properties that reinforce the geometric interpretation of the proposed theory. Moreover, $\bm{M}$ also provides some interesting physical and egineering insights. These are explained below.
%La potencia geométrica $\bm{M}$ y la corriente instantánea $\bm{i}$ cumplen una serie de propiedades matemáticas interesantes y que refuerzan la interpretación geométrica de la descomposición propuesta.
%
\subsection{Orthogonality of $\bm{M}$ components}
The parallel geometric power $M_p$ is a scalar number, while the quadrature geometric power $\bm{M_q}$ is a bivector. Therefore, their internal product is zero and this implies orthogonality:
%es un bivector por lo que es evidente que su producto interno es cero, y por tanto, son ortogonales. 
%The parallel geometric power $(M_p)$ and the quadrature geometric power $(\bm{M_q})$ are orthonormal:
%La potencia geométrica paralela $M_p$ y la potencia geométrica de cuadratura $\bm{M_q}$ son ortogonales
\begin{equation}
	M_p\cdot\bm{M}_q=0
	\label{eq:mp_mq_orthogonal}
\end{equation}
By definition, the norm of the instantaneous geometric power is:
%Además, el módulo de la potencia geométrica vale
%
\begin{equation}
	\|\bm{M}\|=\sqrt{\left<\bm{M}\bm{M}^\dagger\right>_0} 
	\label{eq:modulo_M}
\end{equation}
Therefore, the following relationship can be proven:
\begin{align}
	\|\bm{M}\|^2&=\left<\left(M_p+\bm{M}_q\right)\left(M_p+\bm{M}_q\right)^\dagger \right>_0 \nonumber
	= \|M_p\|^2+\|\bm{M}_q\|^2
	\label{eq:quad_M}
\end{align}
Moreover, the norm of the geometric power also satisfies:
\begin{equation}
    \|\bm{M}\|=\|\bm{u}\|\|\bm{i}\|
\end{equation}
It can be readily demonstrated by:

\begin{equation}
\begin{aligned}
	\|\bm{M}\|
	&=
	\sqrt{\langle\bm{M}^{\dagger}\bm{M}\rangle_0}
	=
	\sqrt{\langle\left(\bm{ui}\right)^{\dagger}\left(\bm{ui}\right)\rangle_0} \\
	&=
	\sqrt{\langle\left(\bm{i}^{\dagger}\bm{u}^{\dagger}\right)\left(\bm{ui}\right)\rangle_0}
	=
	\sqrt{\|\bm{u}\|^2\|\bm{i}\|^2}
	=
	\|\bm{u}\|\|\bm{i}\|
	\end{aligned}
\end{equation}

\subsection{Conservative property for $\bm{M}$}
The instantaneous geometric power $\bm{M}$ is conservative and fulfils the Tellegen's theorem, which means that its sum over across all components in a circuit is zero. This allows for identification of sources and sinks of reactive power, which may lead to allocation of compensation requirements.

\subsection{Sign of the quadrature power $\bm{M}_q$}

The sign of $\bm{M}_q$ provides useful information about the electrical characteristics of a load. Its value is positive for inductors, negative for capacitors,and null for resistors.

\subsection{Averaged value of $\bm{M}_q$}
The averaged value of $\bm{M}_q$  ($\bar{\bm{M}}_q$) is related to energy
storage in inductors and capacitors (in the Budeanu sense)  \cite{jeltsema2015budeanu,willems2010budeanu}.

\subsection{Orthogonality of current components}
As shown in (\ref{eq:current_decomposition_initial}), the instantaneous geometric current $\bm{i}$ can be decomposed into two vectors, $\bm{i_p}$ and $\bm{i_q}$.
%La corriente instantánea $\bm{i}$ se descompone de manera general en dos vectores $\bm{i_p}$ e $\bm{i_q}$ según  (\ref{eq:current_decomposition_initial}).
%
Based on (\ref{eq:mp_mq_orthogonal}), these vectors are orthogonal:
%Se demuestra que ambos vectores son ortogonales 
%
\begin{equation}
	\bm{i}_p\cdot \bm{i}_q=\frac{\bm{u}}{\|\bm{u}\|^2}M_p\cdot\frac{\bm{u}}{\|\bm{u}\|^2}\bm{M}_q=\frac{\|\bm{u}\|^2}{\|\bm{u}\|^2} M_p \cdot \bm{M}_q =0
	\label{eq:ortogonal_corriente}
\end{equation}
Also, it can be proven that all the terms in (\ref{eq:todas_corrientes}) and (\ref{eq:todas_corrientes2}) are orthogonal to each other. This fact establishes a solid framework for current reference in compensation applications like those as passive or active filtering.
\section{Instantaneous Geometric Impedance}
For a linear load, the instantaneous relationship between in-band components of phase voltages and currents can be defined as the phase \textit{instantaneous geometric impedance}. For a given phase $k$, the current and voltage can be represented either in Cartesian or in polar coordinates as~\cite{Montoya2020revisiting}:

\begin{equation}
	\begin{alignedat}{5}
		\bm{u}_k&=\dfrac{1}{\sqrt{2}}(u_k & & \bm{\sigma}_{k}+\bm{\mathcal{H}}\left[u_{k}\right] \bm{\sigma}_{\hat{k}})& & =\dfrac{U}{\sqrt{2}}e^{\varphi_k^v\bm{\sigma}_{k\hat{k}}} & & \bm{\sigma}_{k} \\ 
		\bm{i}_k&=\dfrac{1}{\sqrt{2}}(i_k & & \bm{\sigma}_{k}+\bm{\mathcal{H}}\left[i_{k}\right]  \bm{\sigma}_{\hat{k}})& & =\dfrac{I}{\sqrt{2}}e^{\varphi_k^i\bm{\sigma}_{k\hat{k}}} & & \bm{\sigma}_{k}
	\end{alignedat}
\end{equation}
\noindent where $e^{\varphi_k^v\bm{\sigma}_{k\hat{k}}} $ is a {\em rotor}, which is used to perform a rotation of $\varphi$ degrees to vectors in the Argand plane $\bm{\sigma}_k - \bm{\sigma}_{\smash{\hat{k}}}$. Therefore:
%de forma que la impedancia geométrica instantánea queda definida por
%
\begin{equation}
	\begin{aligned}
		\bm{Z}_k&=\bm{u}_k\bm{i}_k^{-1}=Ue^{\varphi_k^v\bm{\sigma}_{k\smash{\hat{k}}}}\bm{\sigma}_{k}I^{-1}e^{\varphi_k^i\bm{\sigma}_{k\smash{\hat{k}}}}\bm{\sigma}_{k}\\
		&=\frac{U}{I}e^{(\varphi_k^v-\varphi_k^i)\bm{\sigma}_{k\smash{\hat{k}}}} 
		=R_k+X_k\bm{\sigma}_{k\smash{\hat{k}}}=Z_k\angle\varphi_k
	\end{aligned}
	\label{eq:impedance}
\end{equation}
In the above expression, $Z_k=U/I$ is the instantaneous impedance norm, $\varphi_k=\varphi_k^v-\varphi_k^i$ is the instantaneous phase, $R_k=Z_k \cos \varphi_k$ represents the instantaneous resistance and $X_k=Z_k \sin \varphi_k$ is the instantaneous reactance. 
\section{Example I: Single-Phase Circuit}
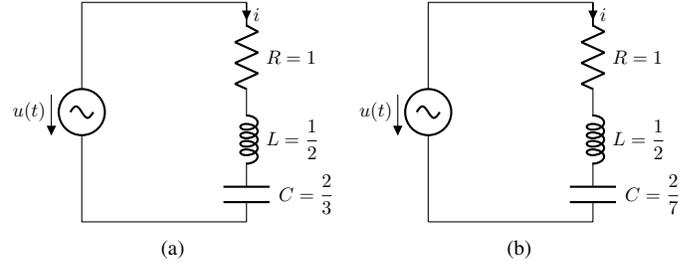
\begin{figure}
	%\centering
	\begin{subfigure}[b]{0.48\columnwidth}
		\begin{circuitikz}[scale=.73,transform shape] \draw
			(0,0) to[sV, v<=$u(t)$] (0,4) -- (2,4)
			-- (3,4)
			(3,0) -- (0,0)
			(3,4) to [R, label=${R=1}$, i>^=$i$] (3,2)
				to [L, label=${L=\dfrac{1}{2}}$] (3,1)
					to [C, label=${C=\dfrac{2}{3}}$] (3,0)
						%	(1.5,4) to[R, european resistor, label=$Y_{cp}$, i>^=$i_{cp}$] (1.5,0)
						;						
						%	\draw (-0.3,2.3) node{$+$};						
			\end{circuitikz}
		  \caption{}
%		\label{fig:RLC_a}
	\end{subfigure} \hfill
	\begin{subfigure}[b]{0.48\columnwidth}
		\begin{circuitikz}[scale=.73,transform shape] \draw
		(0,0) to[sV, v<=$u(t)$] (0,4) -- (2,4)
		-- (3,4)
		(3,0) -- (0,0)
		(3,4) to [R, label=${R=1}$, i>^=$i$] (3,2)
		to [L, label=${L=\dfrac{1}{2}}$] (3,1)
		to [C, label=${C=\dfrac{2}{7}}$] (3,0)
		%	(1.5,4) to[R, european resistor, label=$Y_{cp}$, i>^=$i_{cp}$] (1.5,0)
		;						
		%	\draw (-0.3,2.3) node{$+$};						
	\end{circuitikz}
						  \caption{}
%				\label{fig:RLC_b}
				\end{subfigure}
\caption{The $RLC$ circuits used for Example I.}
\label{fig:RLC1}
\end{figure}

%In the following sections, two cases taken from the literature are solved in the time domain by suing the proposed procedure. 
%
%The first one is a single-phase circuit, while the second one is a three-phase circuit.
%These examples include a single and a three-phase circuit.
%
%\subsection{Single-Phase Circuits}
%
Fig.~\ref{fig:RLC1} shows two simple circuits commonly used in the literature to highlight how traditional power theories fail to provide satisfactory results for apparent power computations.
%
%They are of interest since traditional theory is not able to distinguish them in terms of complex apparent power. 
%
Consider a non-sinusoidal voltage supply such as $u(t)=100\sqrt{2}\left(\sin t + \sin 3t\right)$. The reactive power  (in the Budeanu sense) of each harmonic in the reactive elements are equal, but with opposite signs. Therefore, the total average reactive power is zero in both cases.
This problem has already been solved by using GA in the frequency domain~\cite{montoya2020geometric}. 
It was proven that geometric apparent power components can be clearly identified and, therefore, the geometric apparent power value for each of the circuits in Fig.~\ref{fig:RLC1} is different.
Unfortunately, the use of complex algebra does not allow finding the interaction between voltage and current harmonics of different frequencies. In these previous works, it was shown that circuit (a) can be completely compensated by passive elements, but (b) requires an active compensator to achieve unity power factor.
In this paper, the same circuits will be solved, but in the time domain.
%It proves how it is possible to identify differences at power level through the use of GA.
%
%Therefore, it is possible to distinguish between the two circuits by using the geometric apparent power. 

%The two circuits will be solved in the time domain using GAPoT as formulated in this paper.
%
\subsection{Geometric Apparent Power Calculation}
The geometric voltage and current vectors in the time domain are obtained by using (\ref{eq:volt_curr_vector}):
\begin{equation}
\begin{aligned}
\bm{u}&=\dfrac{1}{\sqrt{2}}(u(t)\bm{\sigma}_1+\bm{\mathcal{H}}\left[u(t)\right]\bm{\sigma}_{\smash{\hat{1}}})\\
\bm{i}&=\dfrac{1}{\sqrt{2}}(i(t)\bm{\sigma}_1+\bm{\mathcal{H}}\left[i(t)\right]\bm{\sigma}_{\smash{\hat{1}}})
\end{aligned}
\end{equation}
The current waveforms for circuits (a) and (b) are:
\begin{align}
i_a
&=
50\left(\sin t + \cos t + \sin 3t - \cos 3t\right) \\
i_b
&=
10\sin t + 30\cos t + 90\sin 3t - 30\cos 3t
\end{align} 
From now on, sub-indexes $a$ and $b$ stand for circuit (a) and (b), respectively. The voltage and currents in the geometric domain are:
\begin{equation*}
\begin{aligned}
\bm{u}
&=
100\left(\sin t + \sin 3t\right)\bm{\sigma}_1+100\left(\cos t + \cos 3t\right)\bm{\sigma}_{\smash{\hat{1}}}
\\
\bm{i}_a
&=
50\left(\sin t + \cos t + \sin 3t - \cos 3t\right)\bm{\sigma}_1\\
&+50\left(\cos t - \sin t + \cos 3t + \sin 3t\right)\bm{\sigma}_{\smash{\hat{1}}}
\\
\bm{i}_b
&=
\left(10\sin t + 30\cos t + 90\sin 3t - 30\cos 3t\right)\bm{\sigma}_1\\
&+\left(10\cos t - 30\sin t + 90\cos 3t + 30\sin 3t\right)\bm{\sigma}_{\smash{\hat{1}}}
\end{aligned}
\end{equation*}
The geometric apparent power can be calculated with (\ref{eq:geom_power_time}), yielding.
\begin{align}
\bm{M}_a
&=
M_{p}^a+\bm{M}_{q}^a=\underbrace{10,000 \left(1 + \sin 2t + \cos 2t \right)}_{M_p^a}
\nonumber \\
\bm{M}_b
&=
M_{p}^b+\bm{M}_{q}^b
=
\underbrace{10,000 + 6,000\sin 2t + 10,000\cos 2t }_{M_p^b}\nonumber \\
&-\underbrace{8,000\sin 2t \bm{\sigma}_{1\smash{\hat{1}}}}_{\bm{M}_q^b} \nonumber
\end{align}
In both cases, the active power is $P=\bar{\bm{M}}_p=10,000$~W.
%The active power is the same in both cases, $P=\bar{\bm{M}}_p/2=10,000$ as explained in Section \ref{sec:current_decomposition}.
%
However, for the quadrature geometric power $\bm{M}_q^a=0$ while $\bm{M}_q^b\neq 0$.
%However, the quadrature geometric power is different, since $\bm{M}_q^a=0$, while $\bm{M}_q^b\neq 0$. 
%
From a practical point of view, this implies that circuit (b) would need active filtering elements in order to achieve unit power factor.
%The above implies that circuit b) must be necessarily compensated by active elements to achieve a unit power factor. 
%
In both circuits, the reactive power (in the Budeanu sense) is $Q=\bar{M}_q=0$. Therefore, it is impossible to calculate the required passive compensation in the time domain.
%Note that in both circuits the reactive power (in the Budeanu sense) is  $Q=\bar{M}_q/2=0$, so it is not feasible to know the eventual passive compensation in the time domain. 
%
In this case, a frequency-domain approach should be used to calculate the contribution of each harmonic (one by one) to the geometric apparent power~\cite{montoya2020geometric}. 
\subsection{Current Decomposition}
\begin{figure}
	\centering
	\begin{subfigure}{\linewidth}
	\centering
		\includegraphics[width=1\columnwidth]{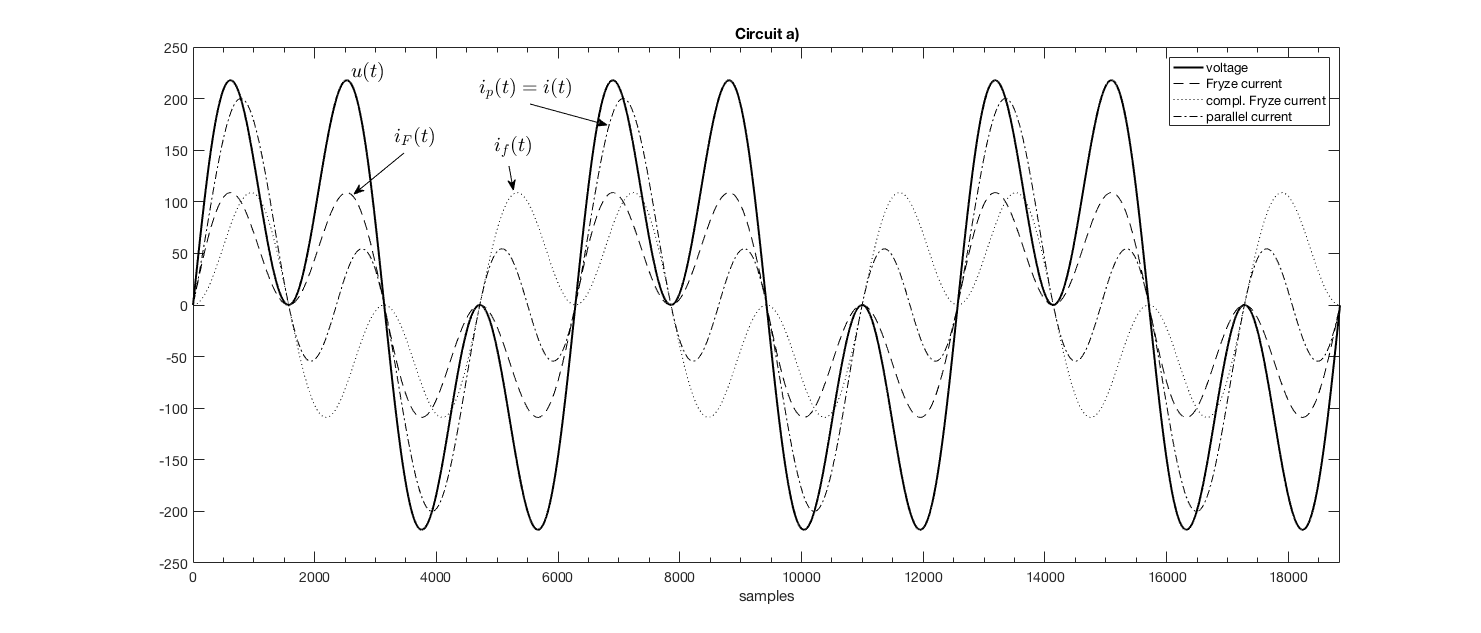}
		\caption{Circuit a}
%		\label{fig:fig1a}
	\end{subfigure}
	\begin{subfigure}{\linewidth}
	\centering
	\includegraphics[width=1\columnwidth]{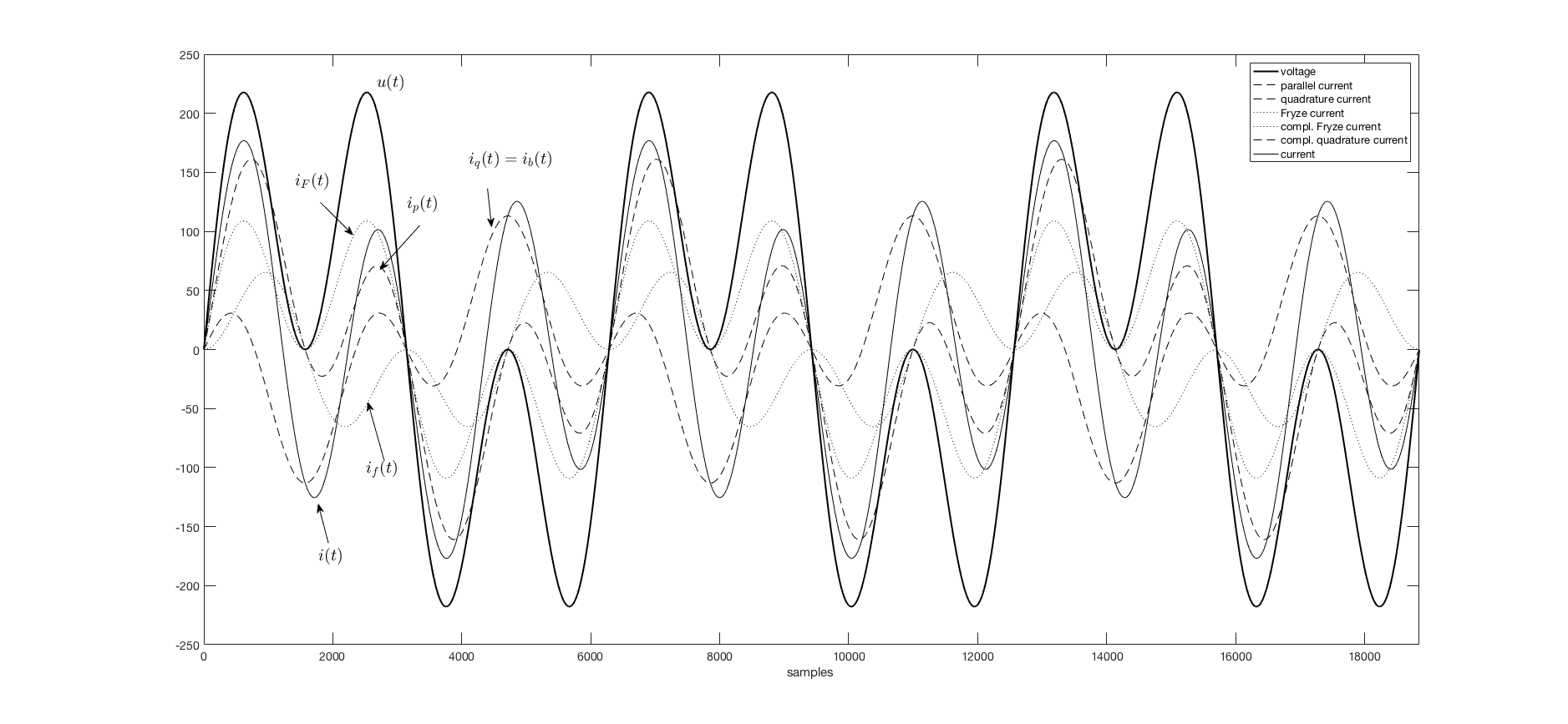}
	\caption{Circuit b}
%	\label{fig:fig1b}
	\end{subfigure}
	\caption{Current in time domain for the circuits in Fig.~\ref{fig:RLC1}.}
	\label{fig:fig1}
\end{figure}

The current can be decomposed by using the expressions (\ref{eq:current_decomposition_initial})--(\ref{eq:todas_corrientes}). 
In order to use them, the inverse of the voltage vector should be calculated:
%It has been taken into account that the inverse of the voltage vector is
%
\begin{equation*}
	\bm{u}^{-1}=\frac{\bm{u}}{\|\bm{u}\|^2}=\frac{\sin t}{100}\bm{\sigma}_1+\frac{\cos 2t}{200\cos t}\bm{\sigma}_{\smash{\hat{1}}}
\end{equation*}
The current components obtained by using this procedure are shown in Table~\ref{tab:RLC_currents_decomposition}. %Note that only the norm of $\bm{\sigma}_1=\bm{\sigma}_{\smash{\hat{1}}}$ is 
%
%\begin{equation}
%	\begin{aligned}
%		\bm{i}_p&=\frac{\bm{u}}{\|\bm{u}\|^2}M_p\\
%		\bm{i}_q&=\frac{\bm{u}}{\|\bm{u}\|^2}\bm{M_q}\\
%		\bm{i}_F&=\frac{\bar{M}_p}{\|\bm{\bar{\bm{u}}}\|^2}\bm{u}\\
%		\bm{i}_f&=\bm{i}_p-\bm{i}_F\\
%		\bm{i}_B&=\frac{\bar{M}_q}{\|\bm{\bar{\bm{u}}}\|^2}\bm{\mathcal{H}}\left[\bm{u}\right]\\
%		\bm{i}_b&=\bm{i}_q-\bm{i}_B
%	\end{aligned}
%\end{equation}
%
\begin{table*}[]
	\centering
	\begin{tabular}{@{}ccllr@{}}
		\toprule
		&  &\multicolumn{2}{c}{\textbf{vector}} & \\ \cmidrule{3-4}
		& \multicolumn{1}{c}{} & \multicolumn{1}{c}{$\bm{\sigma}_1$} & \multicolumn{1}{c}{$\bm{\sigma}_{\smash{\hat{1}}}$} &\multicolumn{1}{r}{$\|\bar{\cdot}\|$}  \\  \cmidrule(lr){3-3} \cmidrule(lr){4-4} \cmidrule(l){5-5}
 		\multirow{2}{*}{$\bm{i}_{p}$}   & a) & $50\sin t + 50\cos t + 50\sin 3t - 50\cos 3t$ & $50\cos t - 50\sin t + 50\cos 3t + 50\sin 3t$    & 100.00\\ [0.2cm]	
		& b) & $50\sin t + 30\cos t + 50\sin 3t - 30\cos 3t$ & $50\cos t - 30\sin t + 50\cos 3t + 30\sin 3t$   & 82.45\\ [0.2cm] \cmidrule{2-5}
	
		\multirow{2}{*}{$\bm{i}_{q}$}   & a) & 0.00 & 0.00   & 0.00\\  
		& b) & $-40\sin t + 40\sin 3t $ &   $-40\cos t + 40\cos 3t $  & 56.56\\ [0.2cm] \cmidrule{2-5}
	
		\multirow{2}{*}{$\bm{i}_{F}$}   & a) & $50\sin t + 50\sin 3t$ & $50\cos t + 50\cos 3t$     & 70.71\\ [0.2cm]
		& b) & $50\sin t + 50\sin 3t$ & $50\cos t + 50\cos 3t$   & 70.71\\ [0.2cm] \cmidrule{2-5}
	
		\multirow{2}{*}{$\bm{i}_{f}$}   & a) & $50\cos t - 50\cos 3t$  & $-50\sin t + 50\sin 3t$     & 70.71\\  
		& b) &  $30\cos t - 30\cos 3t$  &$-30\sin t + 30\sin 3t$& 42.42\\ [0.2cm] \cmidrule{2-5}
	
		\multirow{2}{*}{$\bm{i}_{B}$}   & a) & $0.00$ & $0.00$  &0.00\\  
		& b) & $0.00$ & $0.00$  &0.00\\ [0.2cm] \cmidrule{2-5}
	
		\multirow{2}{*}{$\bm{i}_{b}$}   & a) & $0.00$ & $0.00$  &0.00\\  
		& b) & $-40\sin t + 40\sin 3t $ &   $-40\cos t + 40\cos 3t $  & 56.56\\ \midrule 
	
		\multirow{2}{*}{$\bm{i}$}         & a) & $50\sin t + 50\cos t + 50\sin 3t - 50\cos 3t$ & $50\cos t - 50\sin t + 50\cos 3t + 50\sin 3t$   & 100.00\\  
		& b) & $10\sin t + 30\cos t + 90\sin 3t - 30\cos 3t$ & $10\cos t - 30\sin t + 90\cos 3t + 30\sin 3t$    & 100.00   \\ \bottomrule
	\end{tabular}
	  \caption{Current decomposition for circuit in Figure \ref{fig:RLC1}. The last colum refers to the RMS value of the vector norm.}
	  \vspace{-0.4cm}
	  \label{tab:RLC_currents_decomposition}
\end{table*}
This composition clearly highlights the differences between the two circuits.
%It can be remarked that current decomposition also leads to results that clearly highlights the differences between the two circuits. 
%
First, it can be seen that even though the norm is the same, the waveforms differ among them.
%Although the total current module is the same for both, it is clear that their waveforms differ from each other. 
%
The parallel current $\bm{i}_p$ for circuit (a) is slightly lower compared to that of circuit (b).
%The parallel current $\bm{i}_p$ is slightly lower in circuit b), while the quadrature current $\bm{i}_q$ is zero in circuit a). 
%
The Fryze current $\bm{i}_F$ is the same for both circuits because the active power consumption $P$ is also the same.
%Both circuits exhibit equal Fryze current $\bm{i}_F$ because their active power $P$ is the same. 
%
%This implies that an active compensator should be able to compensate it efficiently until a unity power factor is achieved.
%
In both circuits the current $\bm{i}_B$ is zero. This confirms that it is not possible to compensate their current only with passive elements using the time domain approach. Fig.~\ref{fig:fig1} shows the current components for both circuits. 
Currents in the time domain can be recovered by using (\ref{eq:projection_currents}). 
%%which confirms that it is not possible to compensate them by means of passive elements in the time domain. 
%
\subsection{Geometric Impedance Calculation}
%
%By applying (\ref{eq:projection_currents}) it is possible to retrieve the source currents in the time domain. 
%
%The last column of Table~\ref{tab:RLC_currents_decomposition} shows the module of this current.
%
%
The instantaneous impedance of the circuit can be calculated by using (\ref{eq:impedance}):
\begin{equation*}
	\bm{Z}=\bm{u}\bm{i}^{-1}=\bm{u}\frac{\bm{i}}{\|\bm{i}\|^2}=\frac{\bm{M}}{\|\bm{i}\|^2}
\end{equation*}
\noindent Therefore, for circuits (a) and (b), it follows that
\begin{equation*}
	\begin{aligned}
		\bm{Z}_a&=1+\frac{\cos 2t}{1+\sin 2t} \\
		\bm{Z}_b&=1+\frac{5\cos 2t}{5+3\sin 2t}-\frac{4\sin 2t}{5+3\sin 2t}\bm{\sigma}_{1\smash{\hat{1}}}
	\end{aligned}
\end{equation*}
The two circuits have different behaviours because their instantaneous impedances are also different. It can be seen that circuit (a) does not have instantaneous reactance, while circuit (b) does. Resistive parts also differs.

\section{Example II: Three-Phase Circuit}
\begin{figure}
\centering
\includegraphics[width=0.95\columnwidth]{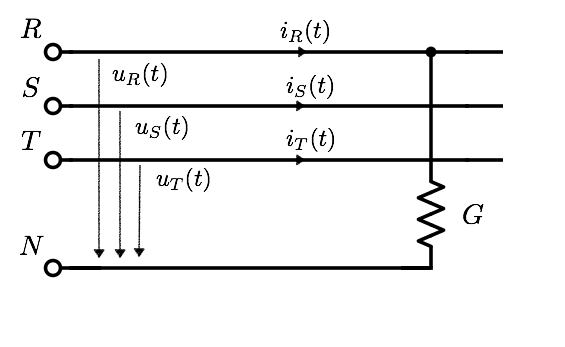}
\vspace{-0.4cm}
\caption{Unbalanced three-phase four wire circuit.}
\label{fig:three_pase}
\end{figure}

Fig.~\ref{fig:three_pase} shows a circuit that has been used by several authors to highlight the weaknesses of the $IRP$ theory~\cite{haley2015limitations,czarnecki2005instantaneous}. 
Only instantaneous  approach is used in the $IRP$ theory. Therefore, some results might be considered senseless or even erroneous.
%This theory does not make use of averaged quantities and therefore obtains results that can be considered uncommon or unnatural.
However, this theory is widely used for instantaneous current compensation.
%This does not preclude some recognition of the merit of instantaneous current compensation. 
%
If the proposed theory in this paper is used, but omitting the HT (\ref{eq:volt_curr_vector}) representation, results would be similar to those obtained with the $IRP$ theory.
%Through GAPoT it is also possible to obtain the same results as the IRP theory just by excluding the Hilbert transform from (\ref{eq:volt_curr_vector}). 
This fact suggests that cross-product theories are a subset of the GA formulation, where the traditional cross vector product is used instead of the exterior product. Unfortunately, the cross product can only be used  in three-dimensional spaces and this limits its applicability. However, the proposed theory can be used for any number of phases.
For a three-phase system, the voltage and current are 
\begin{align}
\bm{u} &= \dfrac{1}{\sqrt{2}}(u_R\bm{\sigma}_1 + u_S\bm{\sigma}_2 + u_T\bm{\sigma}_3) \nonumber \\
\bm{i} &=\dfrac{1}{\sqrt{2}}(i_R\bm{\sigma}_1 +i_S\bm{\sigma}_2 +i_T\bm{\sigma}_3) 
\end{align}
The current decomposition procedure is similar to the previous single-phase circuit by applying (\ref{eq:geom_power_time}) if the instantaneous approach is used.
However, if we apply (\ref{eq:volt_curr_vector}), where the HT is incorporated, the results would be more consistent from the electrical point of view and it would be possible to calculate the minimum current (in the Fryze sense).
%what is expected from an electrical point of view,
%i.e. the minimum current (in the Fryze sense) can be obtained.

In the circuit of Fig.~\ref{fig:three_pase}, the supply voltage is balanced and sinusoidal:
\begin{equation*}
\begin{array}{lll}
u_R(t)=\sqrt{2}\,U\cos \omega t\\  u_S(t)=\sqrt{2}U\,\cos\left(\omega t -120 \right) \\ u_T(t)=\sqrt{2}U\,\cos\left(\omega t +120 \right)
\end{array}
\end{equation*}
Only the current of the phase $R$ is not null. Therefore:
%so that there is only current for the $R$ phase, since the system is unbalanced
%
\vspace{-0.1cm}
\begin{equation*}
i_R(t)=\sqrt{2}GU\cos \omega t
\end{equation*}
\vspace{-0.1cm}
The instantaneous geometric voltage and current vectors are obtained:
\begin{align}
\bm{u} &=U[\cos \omega t\bm{\sigma}_1-\sin \omega t\bm{\sigma}_{\smash{\hat{1}}} +\cos \left(\omega t-120\right)\bm{\sigma}_2 \nonumber \\
&-\sin \left(\omega t-120\right) \bm{\sigma}_{\smash{\hat{2}}} +\cos \left(\omega t+120\right)\bm{\sigma}_3 \nonumber \\
&-\sin \left(\omega t+120\right) \bm{\sigma}_{\smash{\hat{3}}}]
\nonumber  \\[1ex]
\bm{i} &=GU[\cos \omega t\bm{\sigma}_1 -\sin \omega t\bm{\sigma}_{\smash{\hat{1}}}] \nonumber 
\end{align}
\noindent and the geometric apparent is:
\begin{equation*}
\begin{array}{l}
\bm{M}=\bm{ui}=GU^2\left[1-\cos \omega t \cos \left(\omega t - 120\right)\bm{\sigma}_{12} \right. \\
\left.+\cos \omega t \sin \left(\omega t + 120\right)\bm{\sigma}_{1\smash{\hat{2}}} -\cos \omega t \cos \left(\omega t + 120\right)\bm{\sigma}_{13} \right.\\
\left. +\cos \omega t \sin \left(\omega t + 120\right)\bm{\sigma}_{1\smash{\hat{3}}}-\sin \omega t \cos \left(\omega t - 120\right)\bm{\sigma}_{\smash{\hat{1}}2} \right. \\ \left. +\sin \omega t \sin \left(\omega t - 120\right)\bm{\sigma}_{\smash{\hat{1}}\smash{\hat{2}}}  -\sin \omega t \cos \left(\omega t + 120\right)\bm{\sigma}_{\smash{\hat{1}}3} \right. \\
\left. +\sin \omega t \sin \left(\omega t + 120\right)\bm{\sigma}_{\smash{\hat{1}}\smash{\hat{3}}}    \right]
\end{array}
\end{equation*}
In this expression, the parallel power is constant and its value is $M_p=GU^2$. 
This is reasonable since the circuit is purely resistive and consumes an active power equal to $P=GU^2=\bar{M}_p$. 
The other terms are bivectors that are related to the unbalance components of the load. 
Note that $\bm{\sigma}_{1\smash{\hat{1}}}$, $\bm{\sigma}_{2\smash{\hat{2}}}$ and $\bm{\sigma}_{3\smash{\hat{3}}}$ are not present in this expression. 
This means that there is no reactive power in the Budeanu sense. This is also reasonable since there are no inductive nor capacitive elements.
The current decomposition can now be obtained according to (\ref{eq:current_decomposition_initial})-(\ref{eq:todas_corrientes}). 
Considering that $\|\bm{u}\|^2=3U^2$:
\begin{align}
\bm{i}_p&=\frac{\bm{u}}{\|\bm{u}\|^2}M_p=\frac{G}{3}\bm{u} \nonumber \\
&=\frac{GU}{3}[\cos \omega t\bm{\sigma}_1 -\sin \omega t\bm{\sigma}_{\smash{\hat{1}}} +\cos \left(\omega t-120\right)\bm{\sigma}_2 \nonumber \\
&-\sin \left(\omega t-120\right) \bm{\sigma}_{\smash{\hat{2}}} +\cos \left(\omega t+120\right)\bm{\sigma}_3 \nonumber \\
&-\sin \left(\omega t+120\right) \bm{\sigma}_{\smash{\hat{3}}}]\nonumber \\[1ex]
\bm{i}_q&=\frac{\bm{u}}{\|\bm{u}\|^2}\bm{M}_q=\bm{i}-\bm{i}_p= \nonumber \\
&=\frac{GU}{3}[2\cos \omega t\bm{\sigma}_1 -2\sin \omega t\bm{\sigma}_{\smash{\hat{1}}} -\cos \left(\omega t-120\right)\bm{\sigma}_2 \nonumber \\
&+\sin \left(\omega t-120\right) \bm{\sigma}_{\smash{\hat{2}}} -\cos \left(\omega t+120\right)\bm{\sigma}_3 \nonumber \\
&+\sin \left(\omega t+120\right) \bm{\sigma}_{\smash{\hat{3}}}]\nonumber 
\end{align}
%
%		\bm{i}_F&=\frac{\bar{M}_p}{\|\bm{\bar{\bm{u}}}\|^2}\bm{u}\\
%		\bm{i}_f&=\bm{i}_p-\bm{i}_F\\
%		\bm{i}_B&=\frac{\bar{M}_q}{\|\bm{\bar{\bm{u}}}\|^2}\bm{\mathcal{H}}\left[\bm{u}\right]\\
%		\bm{i}_b&=\bm{i}_q-\bm{i}_B
%
For this example, the Fryze current matches the parallel current $(\bm{i}_p=\bm{i}_F)$. Therefore, $\bm{i}_f=0$. 
Also, there is no reactive current (in the Budeanu sense) since $\bar{M}_q=0$, and therefore, $\bm{i}_B=0$. 
It can be seen that $\bm{i}_b=\bm{i}_q$. This means that this current contains all the asymmetrical components, which include both zero sequence current $\bm{i}_0$ and negative sequence current $\bm{i}_{-}$:
\begin{equation*}
\bm{i}_q=\bm{i}_b=\bm{i}_0+\bm{i}_{-}
\end{equation*}
\vspace{-0.4cm}
\begin{align}
\bm{i}_0&=\frac{GU}{3}[\cos \omega t\bm{\sigma}_1 -\sin \omega t\bm{\sigma}_{\smash{\hat{1}}}+\cos \omega t\bm{\sigma}_2 \nonumber \\
&-\sin \omega t\bm{\sigma}_{\smash{\hat{2}}}+\cos \omega t\bm{\sigma}_3 -\sin \omega t\bm{\sigma}_{\smash{\hat{3}}}] \nonumber \\
\bm{i}_{-}&=\frac{GU}{3}[\cos \omega t\bm{\sigma}_1 -\sin \omega t\bm{\sigma}_{\smash{\hat{1}}} +\cos \left(\omega t+120\right)\bm{\sigma}_2 \nonumber \\
&-\sin \left(\omega t+120\right) \bm{\sigma}_{\smash{\hat{2}}} +\cos \left(\omega t-120\right)\bm{\sigma}_3 \\&-\sin \left(\omega t-120\right) \bm{\sigma}_{\smash{\hat{3}}}]
\end{align}

\begin{figure}[!t]
\centering
\includegraphics[width=0.9\columnwidth]{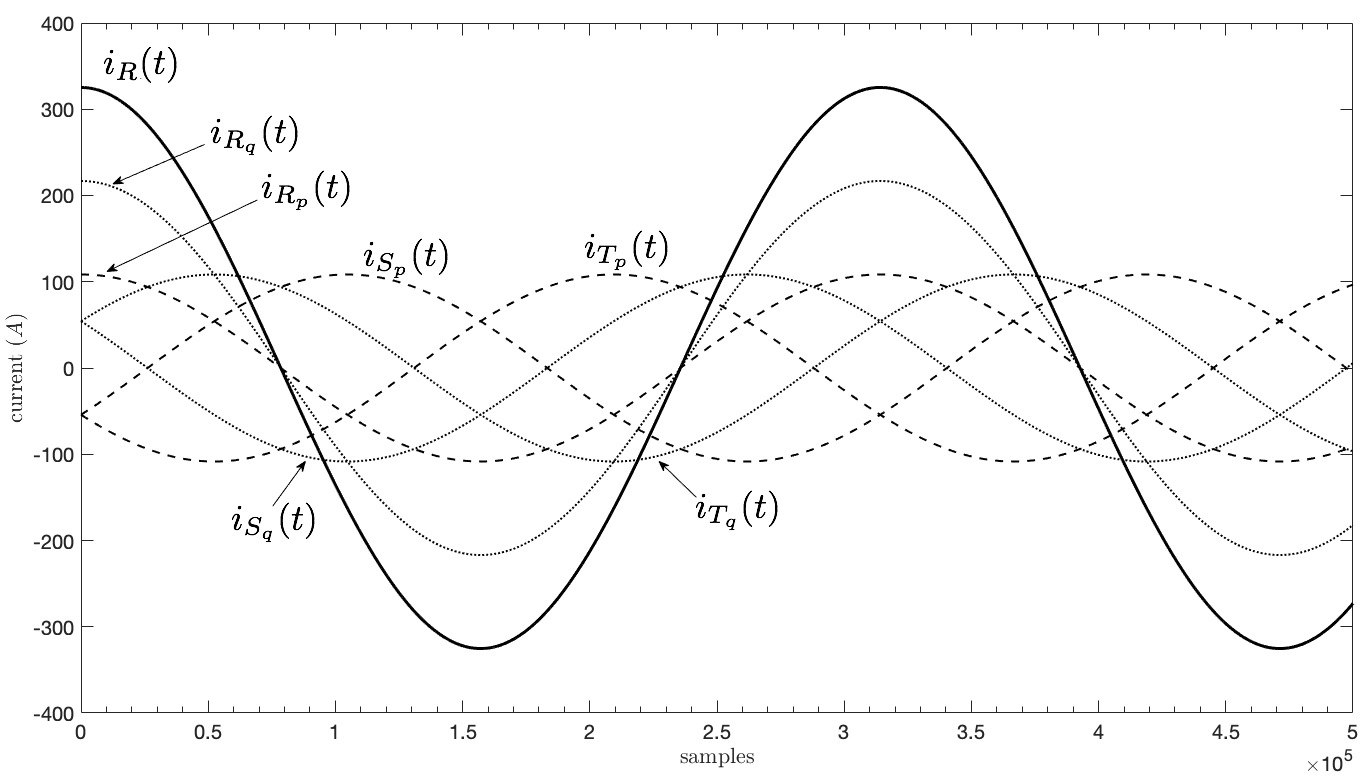}
\caption{Currents decomposition for three-phase circuit.}
\label{fig:three_pase_currents}
\end{figure}

Fig.~\ref{fig:three_pase_currents} shows the current decomposition for $U=230$~V, $\omega=1$~rad/s and $G=1$ Ohm. 
The time domain current is obtained by applying (\ref{eq:projection_currents}) to the geometric vector
\begin{equation*}
i_p(t)=
\begin{bmatrix*}[c]
i_{R_p}(t) \\ i_{S_p}(t) \\ i_{T_p}(t)
\end{bmatrix*}
=\frac{\sqrt{2}GU}{3}
\begin{bmatrix*}[c]
\cos \omega t \\ \cos\left(\omega t -120 \right) \\ \cos\left(\omega t +120 \right)
\end{bmatrix*}
\end{equation*}
\begin{align}
i_q(t)&=
\begin{bmatrix*}[c]
i_{R_q}(t) \\ i_{S_q}(t) \\ i_{T_q}(t)
\end{bmatrix*}
=\frac{\sqrt{2}GU}{3}
\begin{bmatrix*}[c]
2\cos \omega t \\ -\cos\left(\omega t -120 \right) \\ -\cos\left(\omega t +120 \right)
\end{bmatrix*} \nonumber \\
&=\underbrace{\frac{\sqrt{2}GU}{3}
\begin{bmatrix*}[c]
\cos \omega t \\ \cos \omega t \\ \cos \omega t 
\end{bmatrix*}}_{i_0(t)}
+\underbrace{\frac{\sqrt{2}GU}{3}
\begin{bmatrix*}[c]
\cos \omega t \\ \cos\left(\omega t +120 \right) \\ \cos\left(\omega t -120 \right)
\end{bmatrix*}}_{i_{-}(t)} \nonumber 
\end{align}
%
%=\frac{\sqrt{2}GU}{3}\cos \omega t
%=\frac{\sqrt{2}GU}{3}\cos\left(\omega t -120 \right)
%=\frac{\sqrt{2}GU}{3}\cos\left(\omega t +120 \right)
%
\section{Conclusions}
In this paper, power definitions and computations for multi-phase electrical circuits in the time domain have been addressed by using geometric algebra and the Hilbert transform. 
It has been shown that the use of these tools greatly simplify mathematical expressions, leading to a compact formulation. Moreover, it establishes a new framework that can be applied under any supply or load condition.
The proposed formulation can be used in both single- and multi-phase systems and it provides a robust  interpretation that is in good agreement with electrical engineering evidences. 
%
%To the best of our knowledge, no other power theory is able to achieve these results.  
%
Current decomposition can be carried out as the inverse operation. The results highlight the validity of the methodology, which has been verified in electrical circuits previously proposed in the literature. 
Further research will be focus on the application of this theory for calculating power flows in distorted electrical networks with non-linear loads.
%Further research will analyse power flows in non-ideal electrical environments such as distorted distribution systems.
%\section*{References}
 \bibliographystyle{IEEEtran} 
\bibliography{GAPOTt}

\end{document}